\documentclass[fleqn,usenatbib]{mnras}

\usepackage{newtxtext,newtxmath}
\usepackage[T1]{fontenc}

\DeclareRobustCommand{\VAN}[3]{#2}
\let\VANthebibliography\thebibliography
\def\thebibliography{\DeclareRobustCommand{\VAN}[3]{##3}\VANthebibliography}

\usepackage{mathtools, graphicx, hyperref}
\usepackage{amsmath, relsize, xspace, comment, xcolor}

\hypersetup{colorlinks,
		linkcolor = {blue},
		citecolor = {blue},
		urlcolor  = {blue}}

\newcommand{\ipic}{\texttt{iPIC3D}\xspace}

\title[Guide-field ion--electron Relativistic Reconnection]{Guide-Field-mediated Multiscale Instabilities in Relativistic Reconnection}

\author[Deka et al.]
{
    Pranab J. Deka,$^{1}$
    \thanks{E-mail: pranab.deka@kuleuven.be} 
    \thanks{A significant part of this research was carried out during PJD's research visit to Dartmouth College, USA.}
    Fabio Bacchini,$^{1,2}$
    Muni Zhou$^{3}$,
    Camille Granier$^{4,5}$
    \\
    $^{1}$Centre for mathematical Plasma Astrophysics, Department of Mathematics, KU Leuven, 3001 Leuven, Belgium \\
    $^{2}$Royal Belgian Institute for Space Aeronomy, Solar-Terrestrial Centre of Excellence, 1180 Uccle, Belgium \\
    $^{3}$Department of Physics and Astronomy, Dartmouth College, Hanover, NH 03755, USA \\
    $^{4}$Department of Physics, University of Maryland, 7901 Regents Drive, College Park, MD 20742, USA \\
    $^{5}$Canadian Institute for Theoretical Astrophysics, 60 St. George St, Toronto, ON M5S 3H8, Canada
}

\date{Accepted XXX. Received YYY; in original form ZZZ}

\pubyear{\the\year{}}

\begin{document}
\label{firstpage}
\pagerange{\pageref{firstpage}--\pageref{lastpage}}
\maketitle

\begin{abstract}
We investigate magnetic-energy dissipation, current-sheet dynamics, and nonthermal particle acceleration in three-dimensional relativistic reconnection in an electron--ion plasma with a realistic mass ratio. Using particle-in-cell simulations of a double Harris current sheet, we explore a range of ion magnetisations and guide-field strengths to determine how guide fields regulate the overall magnetic energy dissipation. At low magnetisation, $\sigma_i=0.1$, increasing the guide field primarily suppresses reconnection: magnetic-energy dissipation decreases, the growth of tearing modes is weakened, and nonthermal particle acceleration remains inefficient. At higher magnetisations, $\sigma_i=1$ and $\sigma_i=5$, the behaviour changes qualitatively. In the zero-guide-field case, strong drift-kink activity corrugates and broadens the current sheet, inhibiting efficient tearing-mediated reconnection. A weak guide field suppresses this drift-kink-driven disruption, allowing the current sheet to remain laminar and more coherent and thereby enhancing magnetic-energy dissipation. However, once the guide field becomes too strong, reconnection is again suppressed: the onset is delayed, tearing activity weakens, current-sheet compression is reduced, and the system retains a larger fraction of its initial magnetic energy. This non-monotonic behaviour is reflected consistently in magnetic-energy evolution, Fourier analysis of the tearing and kink modes, current-sheet thickness, and nonthermal particle acceleration. The most dissipative cases are not necessarily the zero-guide-field runs, but rather those in which the guide field balances drift-kink suppression without strongly impeding the tearing modes. Our results show that the overall system evolution is controlled not only by the available magnetic energy, but also by the guide-field-regulated morphology and stability of the reconnecting current sheet.
\end{abstract}

\begin{keywords}
magnetic reconnection -- relativistic processes -- plasmas -- instabilities -- acceleration of particles
\end{keywords}

%%%%%%%%%%%%%%%%%%%%%%%%%%%%%%%%%%%%%%%%%%%%%%%%%%

%%%%%%%%%%%%%%%%% BODY OF PAPER %%%%%%%%%%%%%%%%%%

\section{Introduction}

High-energy astrophysical sources, including accreting black holes, relativistic jets, pulsar winds, magnetars, and compact-object magnetospheres, commonly exhibit rapid variability and nonthermal emission. These signatures require an efficient mechanism capable of converting stored electromagnetic energy into plasma heating, nonthermal particle acceleration, and radiation on dynamical timescales. In many of these environments, the plasma is magnetically dominated, so the available free energy resides primarily in global magnetic fields rather than in bulk kinetic motion. This makes magnetic reconnection a natural candidate for powering high-energy flares and nonthermal emission in compact-object systems \citep{Guo24review, Sironi25review}.

Magnetic reconnection occurs when oppositely directed magnetic fields are brought into close proximity and the ideal frozen-in condition breaks down at kinetic scales, allowing magnetic topology to change and magnetic energy to be converted into plasma heating, bulk flows, and nonthermal acceleration. In relativistic regimes, where the magnetisation parameter (the ratio of magnetic field energy density to the plasma rest-mass energy density) $\sigma\gtrsim 1$, reconnection proceeds rapidly and can produce hard particle spectra extending to high energies. These make relativistic reconnection a viable candidate mechanism for accelerating particles to the highest energies and interpreting high-energy emission from blazars, pulsar wind nebulae, black-hole coronae, magnetically arrested accretion flows, and neutron-star magnetospheres \citep{Sironi25review}.

The topological rearrangement of the magnetic field is controlled by electron-scale physics, and the resulting nonthermal distributions cannot be captured self-consistently  by magnetohydrodynamics (MHD) models. Particle-in-cell (PIC) simulations have therefore become the primary tool for studying relativistic reconnection from first principles. Large-scale pair-plasma PIC simulations \citep[e.g.][]{Sironi2014} established that relativistic reconnection can self-consistently generate hard nonthermal particle spectra, with power-law slopes becoming harder at higher magnetisation. They also showed that, in three dimensions (3D), drift-kink activity can dominate the early evolution and delay efficient acceleration, but the late-time system may become tearing/plasmoid dominated and recover nonthermal particle acceleration.

Theoretical work has further connected reconnection-energised particles to the structure of the reconnecting current sheet. In plasmoid-dominated reconnection, \citet{Uzdensky2022} argued that the balance between acceleration by the reconnecting electric field and magnetisation by the reconnected field controls the power-law slope of  the particle energy spectra, whilst trapping in plasmoids regulates the high-energy cutoff. This directly links nonthermal particle acceleration to the evolving current-sheet morphology. Recent studies have shown that the injection of particles into the nonthermal tail is also affected by dimensionality and guide-field physics. In weak-guide-field relativistic reconnection, \citet{French2026} found that the injection energy is set by a gyroradius--current-sheet-thickness criterion rather than by a simple $\gamma_{\rm inj}\sim\sigma$ scaling, and that 3D effects increase $\gamma_{\rm inj}$ and steepen the resulting spectra relative to 2D simulations.

In 2D scenarios, reconnection proceeds exclusively through the tearing instability, which fragments a thin current sheet into magnetic islands and enables efficient energy conversion. However, in 3D, this picture changes qualitatively. Flux ropes can kink, fragment, and lose coherence along the current direction, allowing particles to escape magnetic traps and access additional acceleration channels. A  comparison of 2D and 3D current-sheet evolution by \citet{Werner2021} showed that 3D reconnection is not simply a perturbed version of the 2D plasmoid-chain picture. In moderately magnetised pair plasma, they found that 3D evolution can proceed through a sensitive interplay of tearing, drift-kink deformation, and flux-rope decay, whilst maintaining significant nonthermal particle acceleration \citep{Werner2024, Bacchini2025, Hu2026, Camille26, Velberg25}. 

\citet{Zhang2021} showed that, in weak-guide-field relativistic pair-plasma reconnection, particles with $\gamma\gtrsim 3\sigma$ can escape finite-length flux ropes along the current direction and enter the upstream, where they undergo Speiser-like motion and gain energy nearly linearly in time from the large-scale reconnection electric field. Building on this, \citet{Zhang2023} developed an analytical model for 3D power-law formation in which high-energy particles alternate between a rapidly accelerating free phase in the upstream and a trapped phase in reconnected flux ropes. Because the acceleration time and escape time both scale approximately linearly with particle energy, the free particles approach $\mathrm{d}N/\mathrm{d}\gamma\propto\gamma^{-1}$, while the total particle population approaches an approximately universal $\mathrm{d}N/\mathrm{d}\gamma\propto\gamma^{-2}$ spectrum. Nonrelativistic 3D ion--electron reconnection has shown that flux-rope kink instability can generate field-line chaos in weak-guide-field layers, enabling energetic particles to escape flux ropes, continue Fermi acceleration, and form simultaneous ion and electron power-law distributions \citep{Zhang2021PRL}.

In astrophysical systems, reconnecting layers are not necessarily anti-parallel; they often contain a magnetic-field component along the current direction, i.e., a guide magnetic field. This guide field changes particle orbits, modifies the effective Alfv\'{e}n speed, alters the reconnecting electric field, and affects how energy is partitioned between electrons, ions, fields, and bulk motion. Strong guide fields can slow relativistic reconnection by adding magnetic inertia to the outflow, whilst weak or moderate guide fields may suppress disruptive kink activity without completely quenching tearing. The resulting dependence of energy dissipation and particle acceleration on guide-field strength is therefore not necessarily monotonic.

This is particularly important for ion--electron plasmas. In the semirelativistic regime relevant to black-hole accretion flows and coronae, electrons can be ultrarelativistic whilst ions remain subrelativistic. The two species then have different gyroradii, inertia, and energies. 2D PIC studies of relativistic ion--electron reconnection showed that the energy conversion, particle spectra, and ion--electron energy partition depend sensitively on magnetisation and guide-field strength. \citet{Melzani2014a,Melzani2014b} found that higher magnetisation produces harder particle spectra and larger magnetic-to-kinetic energy conversion, while finite guide fields reduce the total converted energy and favour electron energisation relative to ions.

Recent 3D studies have shown that tearing, drift-kink, and flux-rope kink dynamics can compete in semirelativistic ion--electron current sheets, and that the guide field strongly regulates both the amount of magnetic energy released and the ion--electron energy partition. \citet{Werner2024} found that increasing guide-field strength slows the overall energy conversion, and changes the relative energisation of electrons and ions. More recent simulations extending to higher magnetisation further suggest that three-dimensional effects need not always enhance reconnection: at sufficiently high magnetisation, rapid drift-kink activity can thicken the current sheet, delay tearing, reduce the formation of coherent flux ropes, and lower both magnetic-energy dissipation and nonthermal particle acceleration \citep{Bacchini2025}. These results indicate that the competition between tearing and kink-like dynamics is central to understanding magnetic-energy release in realistic three-dimensional ion--electron reconnection.

In this work, we investigate how guide fields regulate ion--electron relativistic reconnection, and under what conditions do they enhance or suppress magnetic-energy dissipation in 3D Harris-sheet configurations. Using a realistic mass ratio of $m_i/m_e=1836$, we explore a range of ion magnetisation, $\sigma_i$, and guide-field strength relative to the reconnecting field, $B_z/B_0$, to understand how the competition between tearing and kink-like modes changes across parameter space. Combining magnetic-energy evolution, instability mode power, current-sheet morphology, and particle energy spectra, we posit a consistent picture of guide-field-mediated 3D relativistic reconnection. Our results elucidate when guide fields enhance reconnection-driven magnetic energy dissipation and when they suppress it.

This paper is organised as follows. In Sec.~\ref{sec:setup}, we describe the numerical setup, initial conditions, and simulation parameters. In Sec.~\ref{sec:cs_evolve}, we examine the temporal evolution and morphology of the current sheet. In Sec.~\ref{sec:B_dissipation}, we quantify magnetic-energy dissipation as a function of magnetisation and guide-field strength. In Sec.~\ref{sec:tear_dk}, we analyse the growth of tearing and kink modes, and relate these instabilities to the observed dissipation trends. In Sec.~\ref{sec:particles}, we investigate particle energisation and nonthermal acceleration. Finally, in Sec.~\ref{sec:conclude}, we summarise our findings and discuss their broader implications.

%%% ==================================================================================== %%%

\section{Numerical Setup}
\label{sec:setup}

We conduct 3D particle-in-cell (PIC) simulations of semirelativistic ion--electron plasma using the publicly available \ipic code\footnote{\url{https://github.com/Pranab-JD/iPIC3D-CPU-SPACE-CoE}}; \citep{Markidis2010} with the \texttt{RelSIM} algorithm \citep{Bacchini23, Schoeffler25}. Following \citet{Werner2024} and \citet{Bacchini2025}, we initialise a 3D system with a standard Harris sheet in a simulation domain of $L_x \times L_y \times L_z$, with $L_x = L_z = L_y/2 = L_0 \simeq 55.3 \rho_i \sigma_i$, where $\rho_i \equiv m_ic/(q_iB_0)$ is a measure of the upstream Larmor radius, and $\sigma_i \equiv B_0^2/(4\pi m_i c^2 n_i)$ corresponds to the magnetisation of the upstream ions. Here, $m_i$, $q_i$, and $n_i$ correspond to the mass, charge, and average number density of the upstream ions (protons), $c$ is the speed of light, and $B_0$ is the upstream in-plane magnetic field. 

The upstream ion--electron plasma is initialised as a relativistic Maxwell--Jüttner velocity distribution (uniform in space) and with dimensionless temperature $\Theta = k_\mathrm{B} T_{i,0}/(m_ic^2) = (m_e/m_i) \, k_\mathrm{B} T_{e,0}/(m_ec^2)= 0.01$, and $m_e/m_i = 1836$. The relativistic electron skin depth ($\lambda$) in the upstream is given by $\lambda_e = \lambda_i \sqrt{\gamma_e \, m_e/m_i} \approx 0.173 \, \lambda_i$, where $\gamma$ corresponds to the upstream Lorentz factor of each species. The upstream electron magnetisation is given by $\sigma_e = \sigma_i m_i \, \gamma_i/(m_e \, \gamma_e) \approx 33 \, \sigma_i$.

We initialise two current sheets in the $xz$-plane and the reversing magnetic field changes sign along the $y$-direction, in a fully periodic domain. The initial magnetic-field setup respects a Harris equilibrium, given by 
\[ B_x(y) = B_0 \left[1 - \tanh\left(\frac{y-L_y/4}{\delta_\mathrm{CS}}\right) + \tanh\left(\frac{y-3L_y/4}{\delta_\mathrm{CS}}\right)\right], \]
where $\delta_\mathrm{CS} = 0.5 c/\omega_{p, i}$ is the half-thickness of the current sheet and $\omega_{p, i}=c/\lambda_i$ is the ion plasma frequency. Here, the reconnecting outflows extend along the $x$-direction, and 3D instabilities develop in the out-of-plane ($z$) direction. The two reversals of $B_x$ along $y$ are needed to ensure periodicity along the $y$-axis. The current sheets are maintained by a hot drifting ion--electron Maxwellian population, the density of which is $5n_i$, providing the necessary pressure for an initial equilibrium. We add a uniform out-of-plane guide field $B_z$ along the $z$-direction. 

We explore three different ion magnetisation regimes --- $\sigma_i = 0.1$, $1$, and $5$ ($\sigma_e = 3.3$, $33$, and $165$) and seven different guide-field strengths, $B_z/B_0 = 0.0, 0.05, 0.1, 0.25, 0.5, 0.75,$ and $1.0$. Rewriting the characteristic scale of the simulation domain ($L_0$) as a function exclusively of $\sigma_i$, one finds $L_0 \propto \sqrt{\sigma_i}$. The size of our simulation domain for the three different magnetisation parameters, summarised in Table~\ref{tab:resolution}, is based on this relation. The resolution is chosen in such a way that one ion plasma skin depth in the upstream is resolved by 6--7 grid cells. The simulation runs, with a total of eight particles per cell per species, are carried out until the system has reached a quasi-steady state, i.e., the dissipated magnetic energy plateaus. This turns out to be at least $4000 \, \omega_{p, i}^{-1}$ for $\sigma_i = 0.1$ and $2000 \, \omega_{p, i}^{-1}$ for $\sigma_i = 1$ and $5$. In this study, we do not consider any initial perturbation --- reconnection is driven purely by numerical (PIC) noise.

\begin{table}
    \centering
    \begin{tabular}{|c|c|c|c|} 
        \hline
        $\sigma_{i}$ & \multicolumn{2}{c|}{Simulation Domain} & Resolution \\
        & $(c/\omega_{p,i})^3$ & $(c/\omega_{p,e})^3$ & \\
        \hline 
        \hline
        0.1 & $17.5 \times 35 \times 17.5$ 
            & $101 \times 202 \times 101$ 
            & $128 \times 256 \times 128$ \\
        \hline
        1.0 & $55 \times 110 \times 55$ 
            & $318 \times 636 \times 318$ 
            & $384 \times 768 \times 384$ \\
        \hline
        5.0 & $128 \times 256 \times 128$ 
            & $740 \times 1480 \times 740$ 
            & $768 \times 1536 \times 768$ \\
        \hline
    \end{tabular}
    \caption{System sizes (normalised to the ion and electron skin depth) for different magnetisations.}
    \label{tab:resolution}
\end{table}

%%% ==================================================================================== %%%

\section{Current-Sheet Evolution}
\label{sec:cs_evolve}

\begin{figure*}
    \centering
    \includegraphics[width=0.9\linewidth]{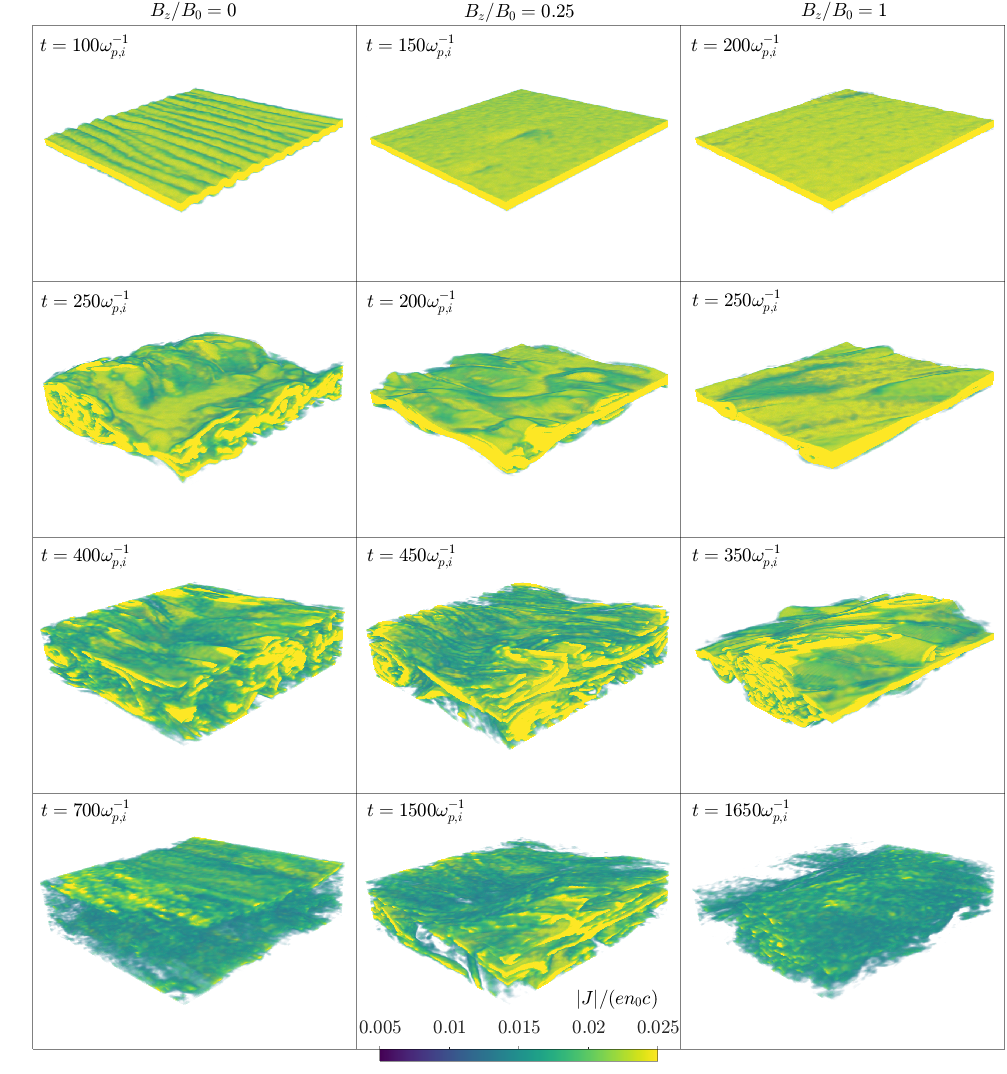} 
    \caption{Time evolution of the current sheet for $\sigma_i=1$ three different guide-field strengths $B_z/B_0$.}
    \label{fig:CS_morphology}
\end{figure*}

We begin by examining the 3D evolution of the current sheet---its morphology is continuously shaped by the competition between tearing, drift-kink, and MHD-kink dynamics. Here, tearing refers to the breakup of the sheet into reconnecting X-points and magnetic islands/flux ropes, drift-kink to a kinetic current-driven corrugation of the sheet along the direction of the current, and MHD-kink to a long-wavelength fluidlike bending or kinking of the entire current sheet. Whether the sheet remains thin and coherent, becomes broadened and corrugated, or relaxes into a more laminar state determines how efficiently magnetic energy is dissipated.

Fig.~\ref{fig:CS_morphology} shows 3D visualisations of the current sheet for $\sigma_i=1$ and three guide-field strengths, $B_z/B_0=0$, $0.25$, and $1$. These snapshots are chosen to illustrate gradual changes in morphology: the initial stage, the onset of reconnection, the strongly dissipative stage, and the late-time state of the layer. The figure highlights the qualitative transition from a strongly corrugated and broadened sheet in the absence of a guide field, to a more coherent but dynamically active sheet at moderate guide field, and a relatively laminar and less active sheet at strong guide field.

The evolution of the current sheet provides a direct way to connect magnetic-energy dissipation (see Sec.~\ref{sec:B_dissipation}) with the competition between tearing, drift-kink, and MHD-kink dynamics (see Sec.~\ref{sec:tear_dk}). The reconnecting layer changes character over time: the system first evolves through a relatively quiescent stage before tearing becomes dynamically important, then enters a stage in which reconnection develops and competes with any pre-existing 3D deformation of the sheet. This is followed by a more dynamically active period of magnetic-energy dissipation, after which the system either continues to reconnect more gradually or relaxes toward a less active state, depending on the guide-field strength.

At high magnetisation, a current sheet is susceptible to the drift-kink instability, leading to the corrugation and broadening of the sheet. As a result, tearing does not grow in an initially thin and coherent layer, but in one that has already been partially puffed up and distorted. Reconnection still develops, but it competes with continuing kink-driven deformation of the layer. This is visible in Fig.~\ref{fig:CS_morphology}, where the $B_z/B_0=0$ case develops strong 3D corrugation at early times. At moderate guide fields, drift-kink broadening is reduced and the tearing modes can grow efficiently. At strong guide fields, the sheet remains significantly smooth, but the onset of reconnection is also impeded.

At low magnetisation, $\sigma_i=0.1$, this behaviour is much weaker or even absent, and the current sheet evolves more as a tearing-dominated layer. A guide field suppresses this early drift-kink deformation and reconnection sheet remains relatively laminar. 

During active reconnection, when magnetic energy is being dissipated most rapidly, morphological differences in the sheet for different guide-field regimes become apparent. For zero or very weak guide fields, the layer remains geometrically disordered, and tearing-mediated energy conversion proceeds in competition with ongoing sheet puffing. At sufficiently high magnetisation, very weak guide fields can still allow drift-kink deformation, hence efficient reconnection requires a guide field strong enough to make the drift-kink modes subdominant.

At moderate guide fields, drift-kink activity may still be present at high magnetisation, but it no longer dominates the current-sheet morphology. The sheet remains thin and more coherent, allowing tearing to remain active over extended times. In addition, long-wavelength oscillations, corresponding to MHD-like kink distortions, along the guide-field direction become visible, especially at late times. Unlike the short-wavelength drift-kink instability, which locally broadens and disrupts the sheet, these longer-wavelength kink-like deformations operate on larger scales and can coexist with efficient reconnection. This regime supports sustained energy dissipation: tearing remains active, MHD-like kink dynamics contributes to large-scale motions of the layer, and the current sheet is not prematurely destroyed by drift-kink broadening.

For strong guide fields, the current sheet remains relatively laminar, as seen in the $B_z/B_0=1$ column of Fig.~\ref{fig:CS_morphology}, and the reconnection dynamics is less vigorous. Magnetic energy is still dissipated, but the active phase is weaker, the high-amplitude current-sheet kinking and the drift-kink corrugation are reduced, and the system more quickly approaches a state in which further energy conversion is inefficient. Here, the guide field over-stabilises the layer: not only does it remove the disruptive drift-kink modes, but it also suppresses the tearing and MHD-like kink activity required for sustained reconnection.

The late-time phase reinforces this interpretation. In the absence of a guide field, the current sheet relaxes into a broad, turbulent, pancake-like structure. Energy dissipation becomes slow and any remaining reconnection proceeds inefficiently within the distorted layer. At moderate guide field, the system remains dynamically active for much longer. Even late in the simulation, reconnection and energy dissipation persist across much of the layer, although at a slower rate than during the most impulsive stage. At strong guide field, the late-time state is significantly inactive: not only has the drift-kink deformation been eliminated, but the tearing and MHD-like kink modes are also suppressed, so the system effectively shuts down.

\begin{figure*}
    \centering
    \includegraphics[width=\linewidth]{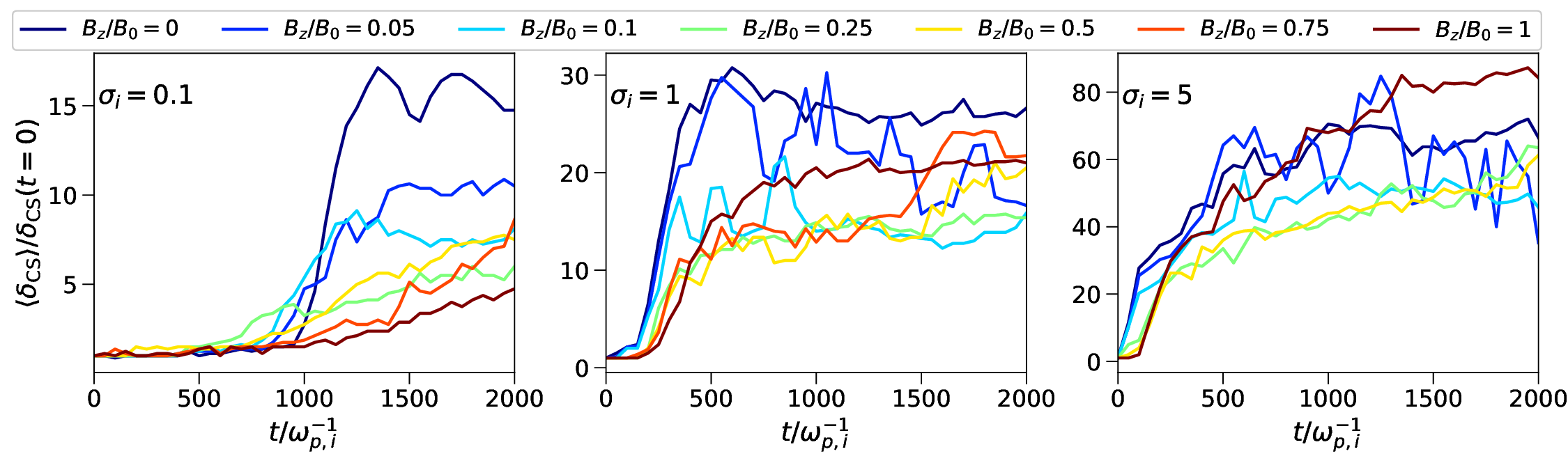} 
    \caption{Time evolution of the current-sheet thickness, $\delta_{\rm CS}/\delta_{\rm CS}(t=0)$, for $\sigma_i=0.1$, $1$, and $5$, and for different guide-field strengths $B_z/B_0$.}
    \label{fig:CS_thickness}
\end{figure*}

Fig.~\ref{fig:CS_thickness} shows the time evolution of the current-sheet thickness, normalised to its initial value, for different guide-field strengths and magnetisations. The thickness serves as a useful proxy to determine which instability is controlling the layer, especially during the early stages. A thin and coherent sheet is more susceptible to tearing and efficient reconnection; an initially broadened, corrugated, or chaotic sheet indicates strong drift-kink activity; and a late-time broadening corresponds to long-wavelength kinking of the sheet.

At low magnetisation, $\sigma_i= 0.1$, the behaviour is relatively straightforward. The zero-guide-field case shows the largest thickness as no stabilising mechanism (guide field) exists to suppress the kinking of the current sheet. Increasing $B_z/B_0$ progressively reduces the broadening of the sheet, whilst simultaneously delaying tearing; nevertheless, the system remains primarily tearing-dominated. This is discussed in further detail in Secs. \ref{sec:B_dissipation} and \ref{sec:tear_dk}.

At $\sigma_i=1$, the thickness evolution becomes more indicative of mode competition. The zero-guide-field case undergoes a rapid early increase in thickness, reaching the largest values amongst the guide-field runs. This suggests that in the absence of a guide field the layer is quickly distorted and broadened by drift-kink activity, which interferes with efficient tearing. Introducing a weak guide field, $B_z/B_0 \sim 0.1$--$0.25$, reduces the broadening by suppressing the drift-kink modes and keeps the layer more coherent. This coincides with the range of guide-field strengths for which magnetic-energy dissipation is enhanced relative to $B_z=0$, indicating that weak guide fields exacerbate reconnection by suppressing kink-driven sheet thickening (see Fig.~\ref{fig:B_evolution_sigma_Bz}). For strong guide fields ($B_z/B_0 \gtrsim 0.5$), the sheet remains thin at early times but the reconnection rate is nevertheless reduced, which implies that the suppression of tearing by the guide field has become more dominant than the suppression of drift-kink leading to current-sheet stabilisation. The thickening of the current sheet for stronger guide fields may be attributed to MHD-like kinking and to the tendency of the hot, overdense sheet to relax toward a broader, lower-free-energy configuration with reduced pressure gradients and magnetic stresses.

We see a similar trend at $\sigma_i=5$. The zero-guide-field case again produces one of the thickest sheets, with rapid early broadening consistent with the strong drift-kink power (discussed in detail in Fig.~\ref{fig:Tear_Kink_S5}). This suggests that drift-kink modes strongly perturb the layer before tearing can efficiently disrupt it. A finite guide field, particularly at $B_z/B_0 \sim 0.1$--$0.25$, limits this broadening and yields a thinner, more coherent reconnecting layer, which correlates with the enhanced magnetic-energy dissipation (shown in Fig.~\ref{fig:Bx_vs_Bz}). At even larger guide fields, the sheet no longer broadens as strongly through kink activity, but reconnection becomes progressively slower because the guide field inhibits thinning by suppressing the tearing modes and reduces the effective outflow speed.

These results suggest that the thickness of the current sheet reflects two competing effects of the guide field. A weak guide field suppresses drift-kink-induced corrugation and broadening, thereby enabling the sheet to remain thin enough for tearing to operate efficiently. However, a strong guide field resists the compression of the sheet and slows down reconnection directly. The non-monotonic dependence of magnetic-energy dissipation on $B_z/B_0$ at $\sigma_i= 1$ and $\sigma_i= 5$ (Fig.~\ref{fig:Bx_vs_Bz}) arises naturally from the balance between these two effects.

Our results are consistent with \citet{Werner2024}, who found that the drift-kink instability thickens the current sheet (in both 2D and 3D), significantly slowing energy conversion. \citet{Bacchini2025} showed that the current sheet becomes increasingly thick and more chaotic at high magnetisation because drift-kink acts earlier and more strongly (in agreement with the results in this paper), suppressing tearing and large-scale flux-rope formation. In contrast, at low magnetisation their current sheets evolve in a 2D-like, tearing-dominated fashion. The recent review by \citet{Sironi25review} emphasises that in 3D the relativistic drift-kink instability ripples and broadens thin current sheets, whilst stronger guide fields slow down reconnection by increasing the effective inertia associated with the frozen-in guide field.

We note the caveat that periodic boundaries, used in this work, may artificially further enhance the thickening of the current sheet by drawing particles from the upstream to the reconnecting layers and continuously replenish the layer with the accelerated particles. Whether allowing for the accelerated particles to leave the simulation domain leads to a significant decrease in the size of the reconnecting layer will be a subject of investigation in our future works. 

%%% ==================================================================================== %%%

\section{Magnetic-Energy Dissipation}
\label{sec:B_dissipation}

Reconnection is initiated by the tearing instability, which fragments a thin current sheet into magnetic islands in 2D and flux ropes in 3D. In our configuration, the growth of the magnetic-field component $B_y$, which is initially set to zero, serves as a diagnostic of the activation of tearing modes, whilst the decay of the reconnecting component $B_x$ quantifies the conversion of magnetic energy into (thermal and nonthermal) particle energy.

\begin{figure*}
    \centering
    \includegraphics[width=\linewidth]{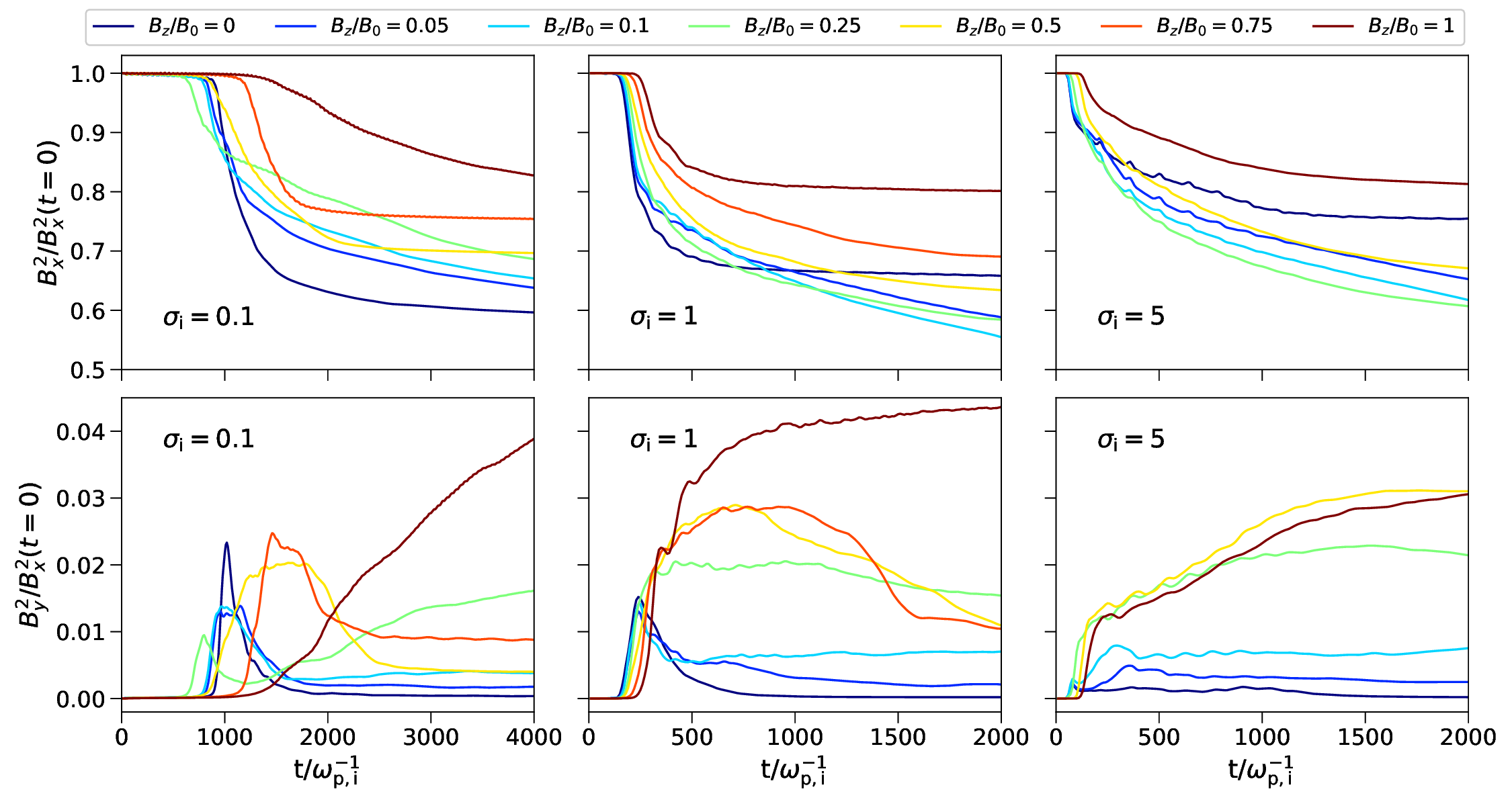}
    \caption{Time evolution of $B_x^2$ (top) and $B_y^2$ (bottom), for the three magnetisations, $\sigma_i=0.1$, $1$, and $5$, and for different guide-field strengths $B_z/B_0$.}
    
    \label{fig:B_evolution_sigma_Bz}
\end{figure*}

In Fig.~\ref{fig:B_evolution_sigma_Bz}, we present the temporal evolution of the magnetic energy in $B_x$ and $B_y$ for the three different magnetisations and a range of guide-field strengths. The effect of the guide field on energy dissipation is not purely monotonic---it reflects competition between tearing and kink-like dynamics, which together regulate the morphology of the current sheet, the details of which are presented in Sec.~\ref{sec:tear_dk}.

At low magnetisation, $\sigma_i=0.1$, the zero-guide-field case undergoes the strongest decay of $B_x^2$ and therefore produces the largest total magnetic-energy dissipation. The dissipation begins after an initial delay and then proceeds rapidly once reconnection becomes active. Increasing $B_z/B_0$ generally reduces the total decrease in $B_x^2$, with the strongest guide-field cases retaining the largest fraction of the initial reconnecting-field energy. This indicates that, at low magnetisation, the guide field primarily acts to stabilise the layer and weaken the reconnection-driven conversion of $B_x$ energy. The corresponding $B_y^2$ evolution is mostly transient for weak and intermediate guide fields: $B_y^2$ rises during the active reconnection phase and then either saturates or decays as the system relaxes. The strong-guide-field case evolves more slowly and shows a delayed growth of $B_y^2$, consistent with a less impulsive and less efficient reconnection phase.

For stronger magnetisations, $\sigma_i=1$ and $\sigma_i=5$, the trend changes qualitatively. A weak but finite guide field can lead to stronger dissipation than the zero-guide-field case, whereas sufficiently strong guide fields reduce the total energy conversion. This non-monotonic behaviour is consistent with the current-sheet morphology discussed above: without a guide field, drift-kink activity can corrugate and broaden the sheet before tearing fully develops, whereas a weak guide field suppresses this disruptive broadening and allows tearing to proceed more efficiently. However, when the guide field becomes too strong, it suppresses not only drift-kink activity but also the tearing and large-scale kink dynamics needed for sustained reconnection.

At $\sigma_i = 1$, weak guide fields ($B_z/B_0 \sim 0.1$--$0.25$) produce the steepest early-time decline in $B_x^2$, indicating a faster reconnection rate compared to both the zero-guide-field and strong-guide-field cases. This enhanced rate leads to a lower final value of $B_x^2$, corresponding to greater total magnetic-energy conversion. In contrast, for strong guide fields ($B_z/B_0 \gtrsim 0.5$), the initial drop in $B_x^2$ is more gradual and the evolution quickly flattens, resulting in a reduced total energy dissipation.

At $\sigma_i = 5$, this effect becomes more pronounced. Although weak-to-intermediate guide fields still yield the largest total dissipation, the rate of energy release is systematically slower than at lower magnetisation. Even in cases with significant dissipation, the decline of $B_x^2$ is more gradual, with no sharp early-time drop comparable to $\sigma_i = 1$. This indicates that reconnection proceeds less impulsively and over longer timescales. For strong guide fields, both the rate and total amount of dissipation are reduced, with $B_x^2$ remaining close to its initial value throughout the evolution.

These results show that increasing magnetisation shifts the optimal guide-field strength for energy dissipation and alters the rate of reconnection. Whilst moderate guide fields can enhance both the rate and total energy release at $\sigma_i = 1$, at higher magnetisation ($\sigma_i = 5$) the system exhibits intrinsically slower energy conversion, even in the most favourable guide-field regime.

\begin{figure}
    \centering
    \includegraphics[width=\columnwidth]{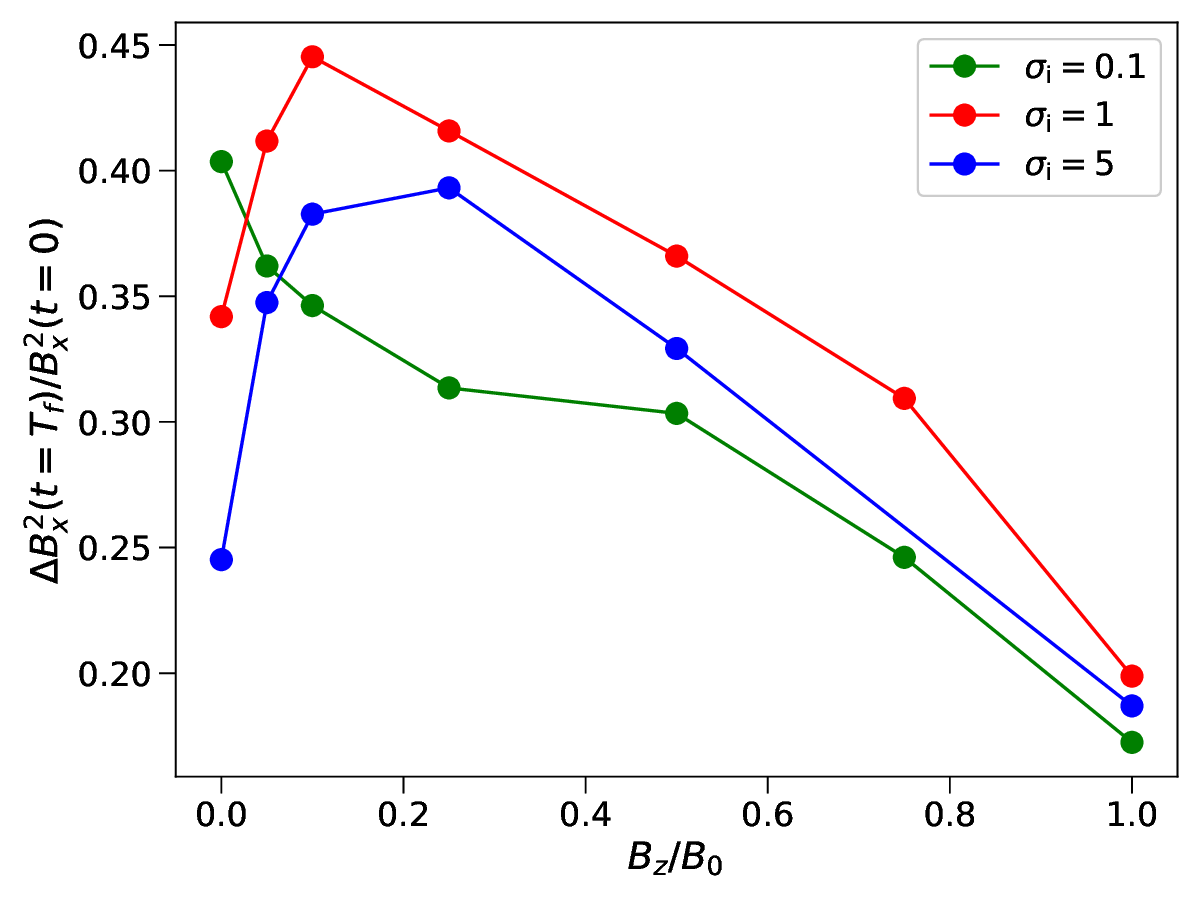}
    \caption{Fraction of reconnecting magnetic energy dissipated by the end of the simulation, as a function of guide-field strength $B_z/B_0$, for $\sigma_i=0.1$, $1$, and $5$.}
    \label{fig:Bx_vs_Bz}
\end{figure}

The non-monotonic magnetic-energy-dissipation trend as a function of guide-field strength is illustrated in Fig.~\ref{fig:Bx_vs_Bz}. For $\sigma_i= 0.1$, magnetic-energy dissipation decreases approximately monotonically with increasing guide-field strength. In contrast, for $\sigma_i= 1$ and $\sigma_i= 5$, the dissipation peaks at weak to intermediate guide fields, indicating an optimal regime in which the guide field is strong enough to mitigate the drift-kink activity, but not strong enough to significantly suppress reconnection or MHD-kink dynamics. A similar result has been reported by \citet{Werner2024}, who found that for $\sigma_i=0.5$ a weak guide field can lead to slightly enhanced magnetic-energy release compared to the zero-guide-field case. However, they caution that this enhancement is modest and may not represent a robust systematic effect, as it is based on a highly limited sampling (in their case) and may be influenced by the inherent nonlinear evolution.

%%% ==================================================================================== %%%

\section{Tearing vs. Kink Modes}
\label{sec:tear_dk}

To better understand how magnetic-energy dissipation depends on guide-field strength and magnetisation, we analyse the temporal evolution of tearing and kink-like perturbations using one-dimensional Fast Fourier Transforms (FFTs) of the magnetic-field components. In what follows, the subscript on the FFT denotes the spatial direction along which the transform is computed. Thus, $\mathrm{FFT}_x(B_y)$ denotes the Fourier transform of $B_y$ along the $x$-direction, whilst $\mathrm{FFT}_z(B_x)$ denotes the Fourier transform of $B_x$ along the out-of-plane ($z$) direction. The angular brackets denote spatial averaging over the remaining direction indicated by the subscript: $\langle |\mathrm{FFT}_x(B_y)|^2 \rangle_z$ is the Fourier power in $B_y$ after transforming along $x$, averaged over $z$ and $\langle |\mathrm{FFT}_z(B_x)|^2 \rangle_x$ is the Fourier power in $B_x$ after transforming along $z$, averaged over $x$.

\begin{figure*}
    \centering
    \includegraphics[width=0.9\linewidth]{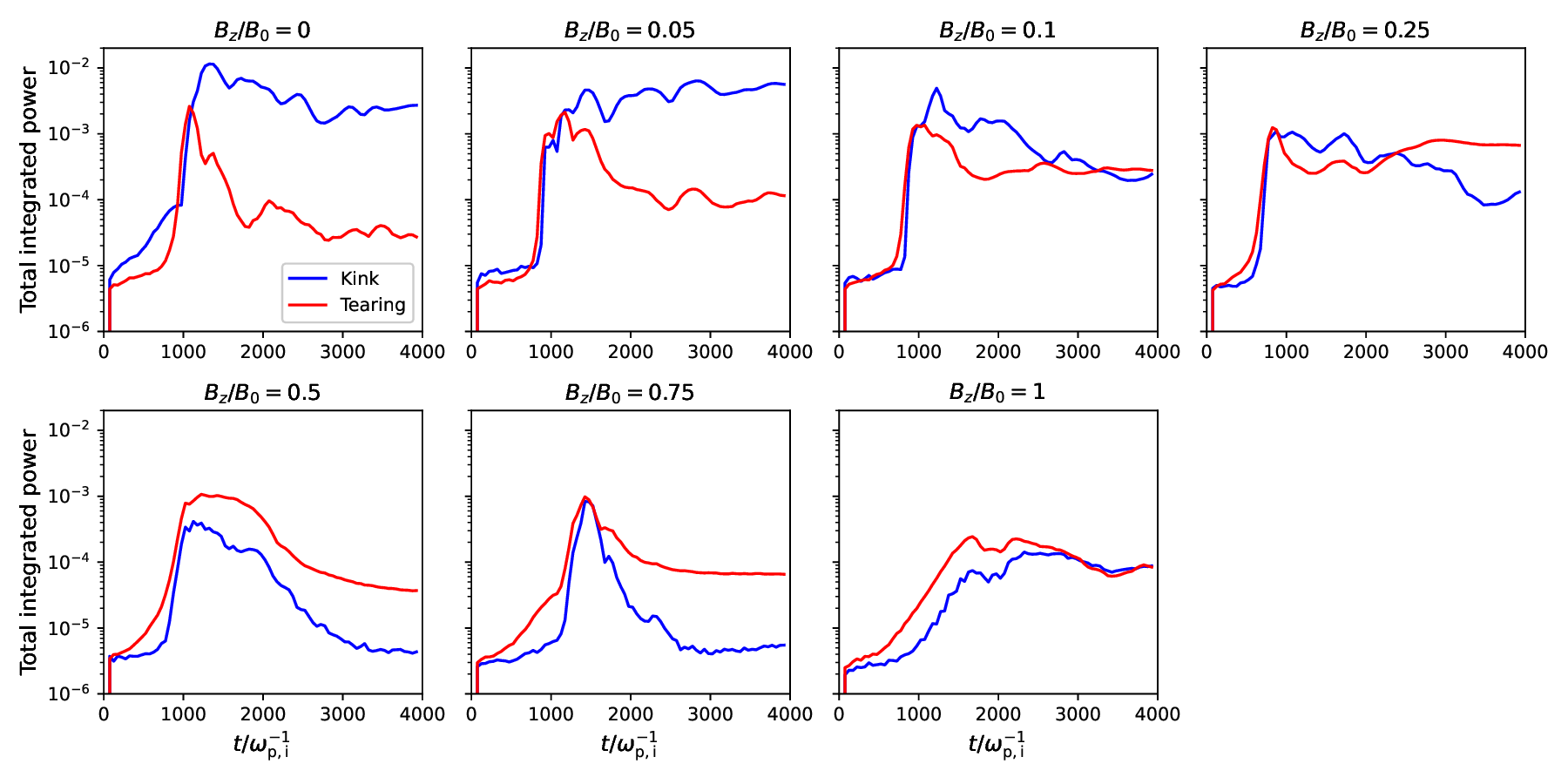}
    \caption{Time evolution of the Fourier power integrated over tearing and kink-like modes for $\sigma_i=0.1$, shown for different guide-field strengths $B_z/B_0$. The red curve denotes the tearing-mode power and the blue curve denotes the kink-mode power, both binned over intervals of $50\,\omega_{p, i}^{-1}$.}
    \label{fig:Tear_Kink_S0.1}
\end{figure*}

\begin{figure*}
    \centering
    \includegraphics[width=0.9\linewidth]{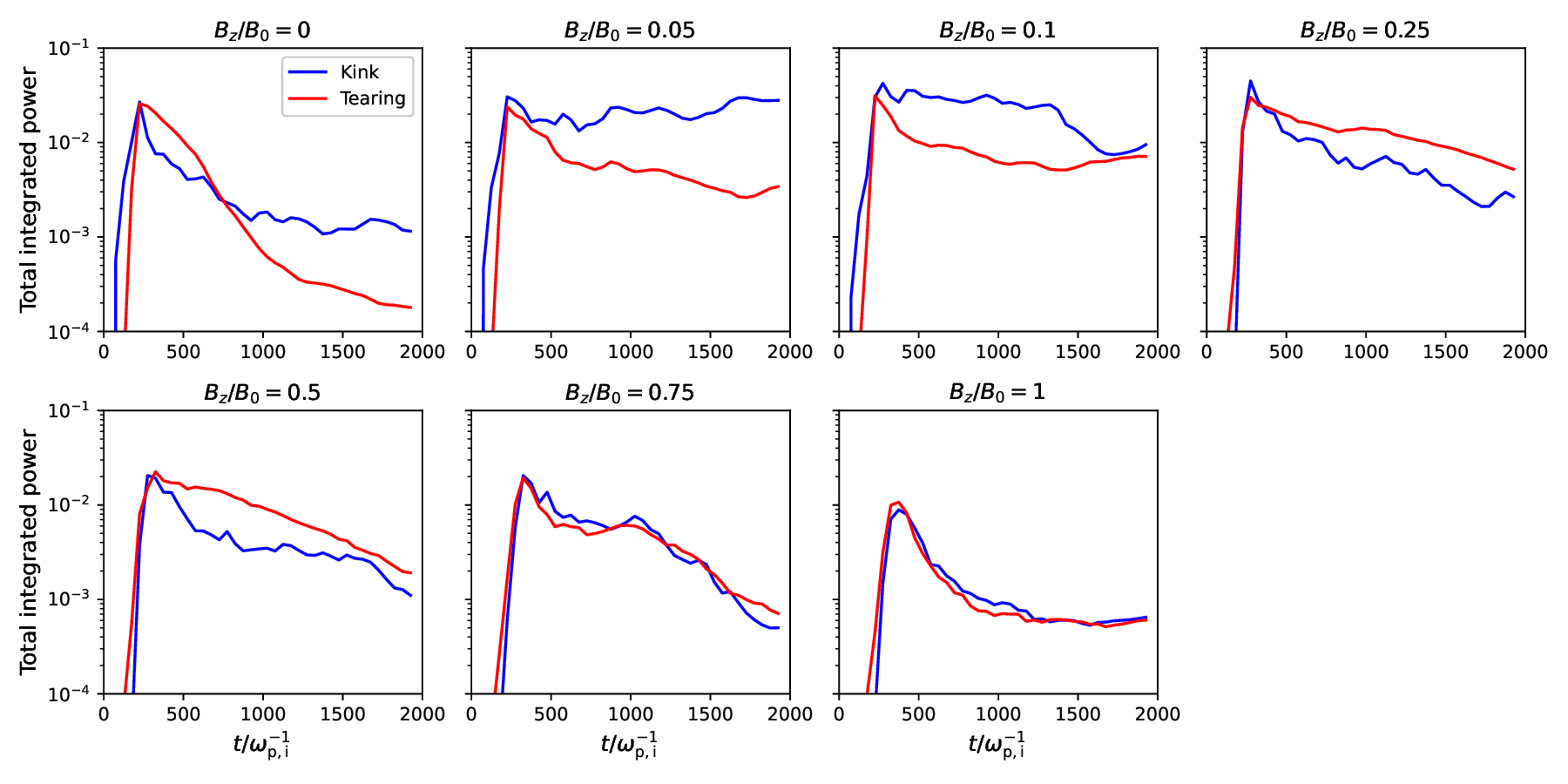}
    \caption{Same as above, but for $\sigma_i = 1$.}
    \label{fig:Tear_Kink_S1}
\end{figure*}

\begin{figure*}
    \centering
    \includegraphics[width=0.8\linewidth]{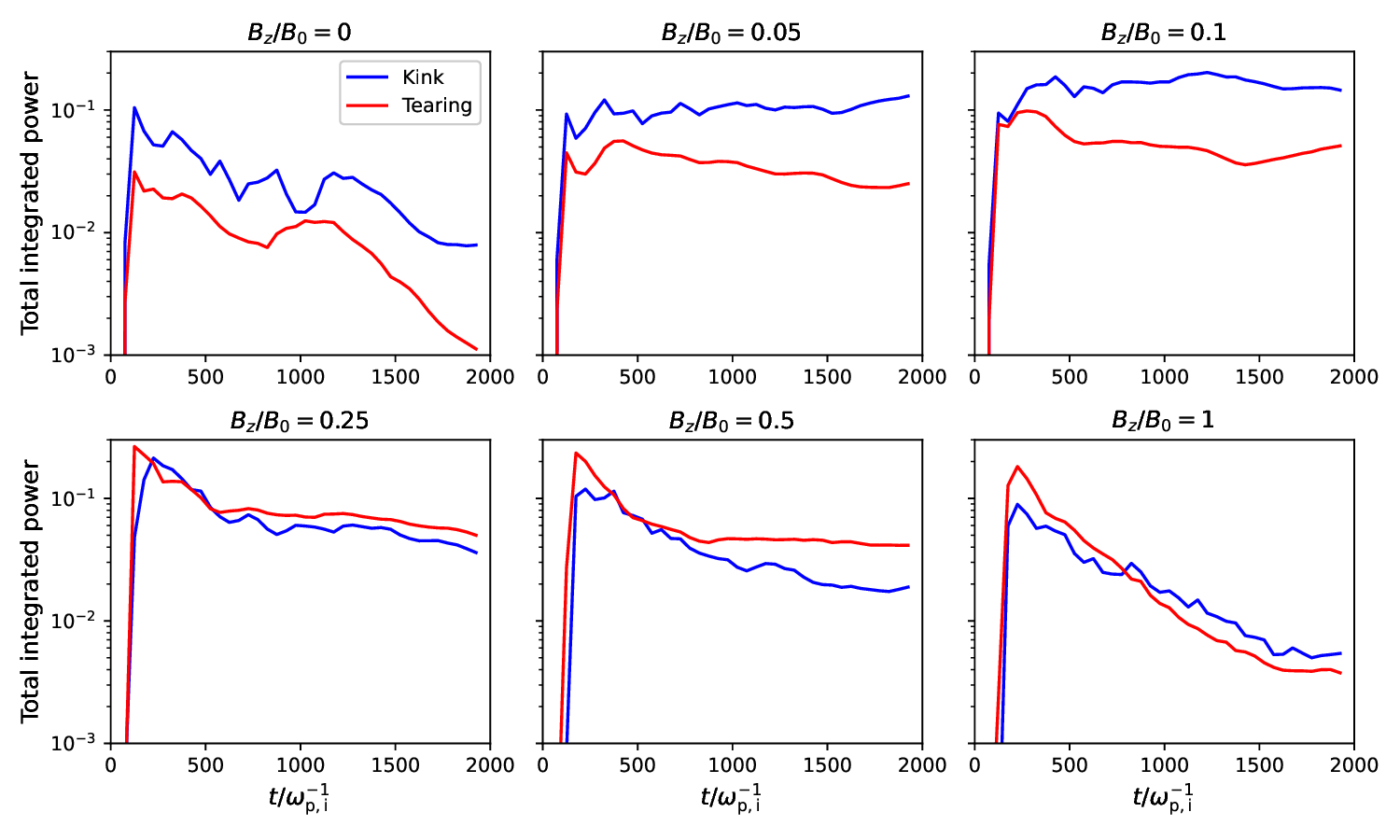}
    \caption{Same as above, but for $\sigma_i = 5$.}
    \label{fig:Tear_Kink_S5}
\end{figure*}

We use $\langle |\mathrm{FFT}_x(B_y)|^2 \rangle_z$ as a proxy for tearing activity because $B_y$ is generated self-consistently by the reconnected topology of the initially antiparallel field. In 2D, this component is associated with magnetic-island formation and in 3D it traces the corresponding flux-rope morphology. Its growth provides a direct measure of tearing-driven restructuring of the current sheet. Here, $x$ is the direction of the reversing, reconnecting magnetic field, and the growth of $B_y$ fluctuations along this direction tracks the formation of X-points, magnetic islands in 2D, and flux ropes in 3D. We use $\langle |\mathrm{FFT}_z(B_x)|^2 \rangle_x$ as a proxy for kink-like modes, since variations of $B_x$ along the out-of-plane direction quantify corrugation and bending of the current sheet. In Figs.~\ref{fig:Tear_Kink_S0.1}, \ref{fig:Tear_Kink_S1}, and \ref{fig:Tear_Kink_S5}, we show the total integrated Fourier power in the tearing-like diagnostic (red lines) and kink-like diagnostic (blue lines) as a function of time. These integrated powers provide a compact measure of which class of instability dominates the current-sheet evolution at a given time.

At $\sigma_i = 0.1$, the drift-kink mode is essentially absent, so the measured kink power is more plausibly produced by an MHD-like kink of the tearing-generated flux ropes. In this magnetisation regime, tearing first drives reconnection and flux-rope formation, after which the flux ropes can undergo kink-like deformation. This distinction is important because the total integrated kink power does not distinguish drift-kink corrugation from MHD-like kinking of the flux ropes. With increasing guide-field strength, tearing remains the primary driver of the evolution at weak and intermediate guide fields, whilst the kink-like deformation follows the tearing-mediated formation of flux ropes. At larger guide fields ($B_z/B_0 \gtrsim 0.5$), both tearing and the kink-like modes are suppressed, producing delayed growth and a more gradual evolution consistent with reduced magnetic-energy dissipation.

At $\sigma_i = 1$, tearing and kink modes compete strongly at weak guide fields, but their relative importance depends on their onset times. In the zero-guide-field case, the kink mode grows earlier and initially dictates the evolution by distorting the current sheet, whilst tearing develops shortly after and reaches comparable or higher power at later times. Introducing a weak guide field ($B_z/B_0 \sim 0.05$--$0.1$) delays and weakens the early dominance of kink modes, allowing tearing to emerge more cleanly on similar timescales. This regime corresponds to the enhanced magnetic-energy dissipation. For intermediate guide fields ($B_z/B_0 \sim 0.25$), tearing increasingly governs the evolution following its onset, with reduced influence from kink-like activity. At larger guide fields ($B_z/B_0 \gtrsim 0.5$), kink modes are strongly suppressed at all times, and tearing dominates the dynamics, although with a smoother (smaller growth rate) and less impulsive evolution.

At $\sigma_i = 5$, kink activity dominates early evolution in the absence of a guide field due to its rapid onset. The kink mode grows on the shortest timescales and reaches a strong peak in power, indicating that the current sheet is quickly distorted and thickened before tearing can fully develop. With the introduction of a weak guide field ($B_z/B_0 \sim 0.1$--$0.25$), the onset of kink modes is delayed and their amplitude reduced (similar to the previous two cases), allowing tearing to develop more effectively and play a larger role in the evolution. For stronger guide fields ($B_z/B_0 \gtrsim 0.5$), kink activity is strongly suppressed from the outset, and tearing clearly dominates the dynamics. However, despite this dominance, the overall evolution becomes more gradual, reflecting the reduced efficiency of reconnection in strongly magnetised guide-field configurations.

% Drift-kink modes are known to broaden and corrugate the current sheet, thereby inhibiting efficient reconnection in the absence of a guide field. Increasing the guide field suppresses these kink-like distortions and promotes the formation of more coherent reconnecting structures, consistent with previous studies \citep{werneruzdensky2021, Hu2026, Camille26}. Our results extend this picture by showing that the balance between tearing and kink activity depends sensitively on magnetisation: the transition from kink-influenced to tearing-dominated dynamics shifts toward weaker guide fields as $\sigma_i$ increases. 

\subsection*{Drift-kink vs. MHD-kink}
\begin{figure*}
    \centering
    \includegraphics[width=0.9\linewidth]{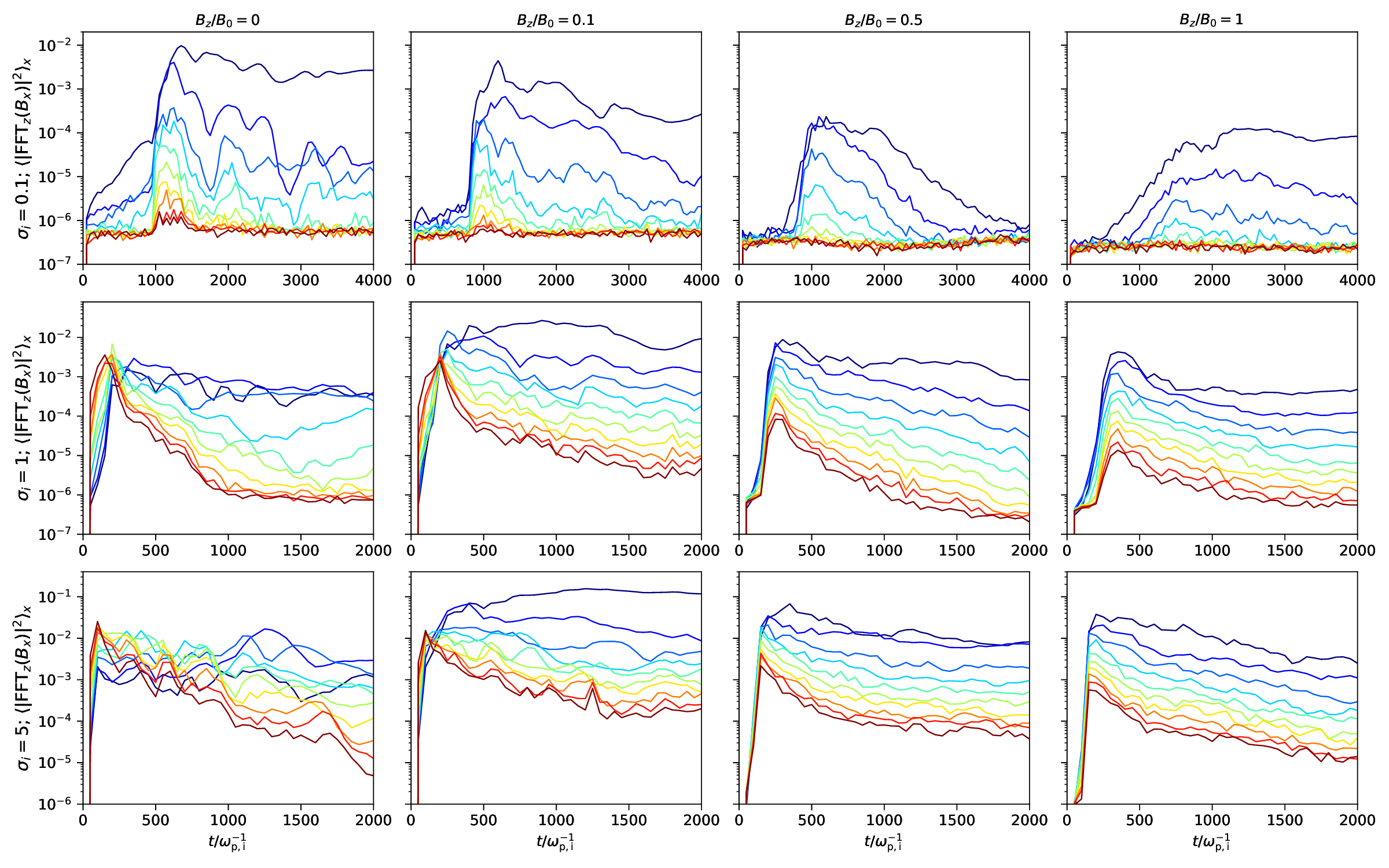}
    \caption{Time evolution of the spectral power of the first ten kink-like modes for $\sigma_i = 0.1$ (top), $\sigma_i = 1$ (middle), and $\sigma_i = 5$ (bottom) and varying guide field strengths. Blue to red lines indicate increasing $k$, from lowest (large-scale) to highest (small-scale) modes.}
     \label{fig:kink_modes}
\end{figure*}

To further disentangle the role of 3D instabilities, we examine the evolution of kink-like fluctuations by decomposing the power spectrum into different wavenumber ranges (Fig.~\ref{fig:kink_modes}). This reveals a clear separation between short-wavelength drift-kink modes (large $k$) and long-wavelength magnetohydrodynamic (MHD) kink modes (small $k$), which exhibit distinct behaviour.

\begin{figure*}
    \centering
    \includegraphics[width=0.9\linewidth]{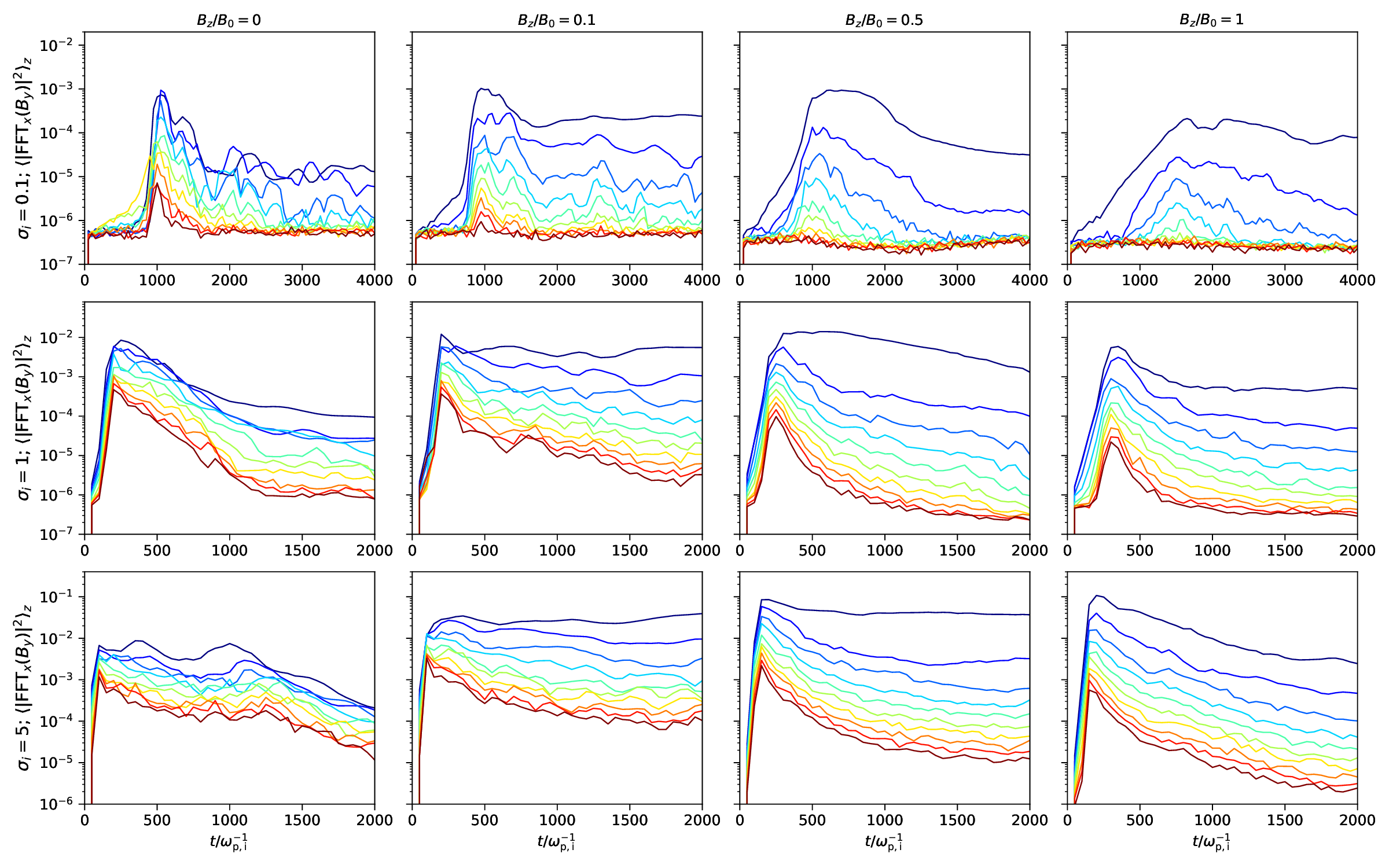}
    \caption{Same as above, but for tearing modes.}
     \label{fig:tear_modes}
\end{figure*}

With an increase in the upstream magnetisation, the drift-kink instability (DKI) exhibits a stronger growth rate than its long-wavelength counterparts. This is seen in the early-time growth of power in the high-$k$ modes (reddish curves), typically peaking at $t \sim 50$--$300\,\omega_{p,i}^{-1}$. The rapid onset of these short-wavelength modes indicates that DKI is the primary mechanism responsible for the initial corrugation and broadening of the current sheet. This early-time distortion disrupts the formation of coherent X-points and delays the onset of efficient reconnection.

At later times, power shifts toward larger spatial scales, with the growth of low-$k$ modes (bluish curves), corresponding to MHD-like kink distortions. These modes evolve more gradually and dominate the late-time structure of the current sheet, especially in cases with weak or moderate guide field. Unlike the DKI, which acts locally and rapidly, the MHD kink operates on global scales and contributes to large-scale kinking (comparable to the domain size) of the current sheet.

The relative importance of these two modes depends strongly on the guide-field strength. In the absence of a guide field, both DKI and MHD kink modes reach large amplitudes, leading to strong multi-scale distortion of the current sheet---on small scales at early times and on large scales at late times. With an increase in the guide-field strength, the high-$k$ DKI modes are preferentially suppressed, resulting in a delayed and weaker early-time peak. This allows for a transition toward a more coherent, tearing-dominated evolution.

At sufficiently large guide fields ($B_z/B_0 \gtrsim 0.5$), kink activity is strongly suppressed throughout the wave spectrum. However, despite the reduction of kink-induced distortions, the overall reconnection dynamics becomes less efficient, consistent with the reduced magnetic-energy dissipation observed at a strong guide field.

These results illustrate that the impact of kink activity on reconnection is intrinsically multiscale: the early-time DKI sets the initial thickness and structure of the current sheet, whilst the later-time MHD kink governs its large-scale deformation. The suppression of these modes by a guide field therefore operates first at small scales and subsequently at large scales, providing a unified explanation for the non-monotonic dependence of reconnection efficiency on guide-field strength.

%%% ==================================================================================== %%%

\section{Nonthermal Particle Acceleration}
\label{sec:particles}

\begin{figure*}
    \centering
    \includegraphics[width=\linewidth]{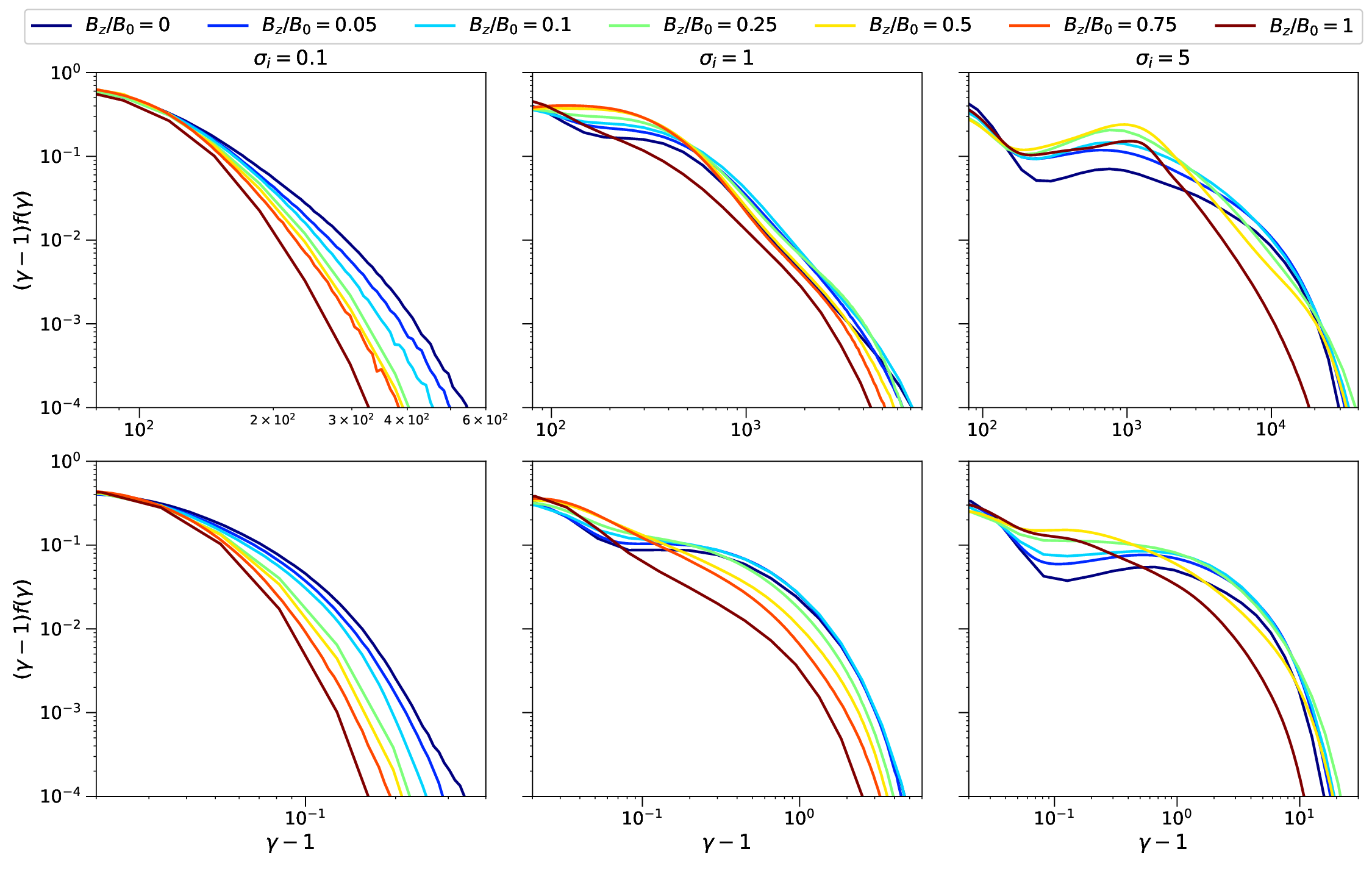}
    \caption{Particle energy distributions at the end of the run for electrons (top row) and ions (bottom row) for $\sigma_i=0.1$, $1$, and $5$, and for different guide-field strengths $B_z/B_0$.}
    \label{fig:PED_sigma_Bz}
\end{figure*}

\begin{figure*}
    \centering
    \includegraphics[width=\linewidth]{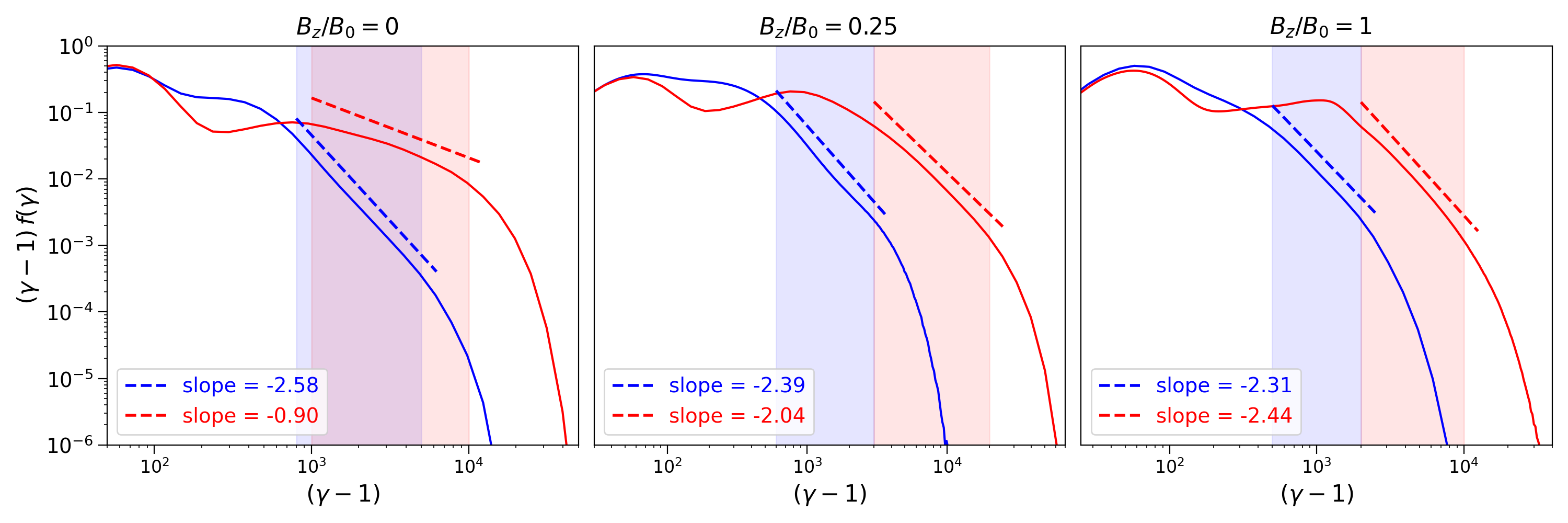}
    \caption{Power-law fits to the nonthermal tails of the final electron energy spectra, $(\gamma-1)f(\gamma)$, for guide-field strengths $B_z/B_0=0$, $0.25$, and $1$. Blue curves correspond to $\sigma_i=1$ and red curves to $\sigma_i=5$; the dashed lines show the fitted slopes over the shaded fitting intervals.}
    \label{fig:EED_sigma_Bz}
\end{figure*}

Fig.~\ref{fig:PED_sigma_Bz} shows the final electron and ion energy spectra for different magnetisations and guide-field strengths, whilst Fig.~\ref{fig:EED_sigma_Bz} illustrates representative fits to the nonthermal part of the electron distributions. Throughout this section, we use the energy-weighted spectrum $(\gamma-1)f(\gamma)$, so that straight segments in log--log space correspond to nonthermal power-law-like intervals in the particle distribution.

At low magnetisation, $\sigma_i=0.1$, both electron and ion spectra remain close to thermal distributions, with only modest broadening relative to the initial state. Increasing $B_z/B_0$ shifts the spectra toward lower maximum energies, consistent with the reduced magnetic-energy dissipation and weaker reconnection dynamics discussed in Sec.~\ref{sec:B_dissipation}. Thus, at low magnetisation, the guide field primarily suppresses reconnection rather than enabling more efficient particle acceleration.

At higher magnetisations ($\sigma_i=1, 5$), the particle spectra show clear suprathermal and nonthermal components. This is most evident for electrons, which develop extended high-energy tails over a range of guide-field strengths. Ions also show suprathermal broadening and, in some cases, nonthermal-like extensions, but these tails are somewhat less pronounced.

The guide-field dependence of the particle spectra is consistent with the non-monotonic magnetic-energy dissipation---weak-to-intermediate guide fields produce the broadest high-energy tails, whilst the strongest guide-field cases show steeper spectra and reduced high-energy cutoffs. In the absence of a guide field, drift-kink activity corrugates and broadens the current sheet, reducing the coherence of the reconnecting layer. A weak guide field partially suppresses drift-kink-driven broadening, allowing a thinner and more coherent current sheet to persist. This favours tearing-mediated reconnection and enables particles to experience a more effective reconnection electric field, whilst also increasing the time they spend in energising regions. The mode decomposition discussed in Sec.~\ref{sec:tear_dk} supports this interpretation: short-wavelength drift-kink modes dominate the early-time distortion of the current sheet, whereas longer-wavelength MHD-like kink distortions contribute to its larger-scale kinking at later times. The particle spectra are therefore controlled not only by the total amount of magnetic energy dissipated, but also by the morphological evolution of the reconnecting layer. If the sheet becomes too broad or chaotic, the acceleration regions become less coherent; if the guide field is too strong, tearing and the reconnection electric field are suppressed, reducing both the energisation time and the high-energy cutoff. The guide field therefore acts as a regulator: too little guide field permits disruptive 3D corrugation, whereas too much guide field suppresses the energising reconnection dynamics. The most efficient acceleration occurs in the intermediate regime, where drift-kink broadening is limited but tearing-driven reconnection remains active.

The representative fits in Fig.~\ref{fig:EED_sigma_Bz} quantify the extension of the nonthermal part of the electron spectra. In selected cases, we fit
\[ (\gamma-1)f(\gamma) \propto (\gamma-1)^{-s} \]
over the shaded intervals. These fits should be interpreted as evidence for the nonthermal component rather than a universal single power law. In the present simulations, the system size and magnetisation are not large enough to produce tails extending over many decades in $\gamma$. Therefore, we focus on robust trends: the relative extension of the nonthermal component and the location of the high-energy cutoff. Both diagnostics show that weak-to-intermediate guide fields can enhance particle acceleration, whereas strong guide fields shorten the tail and reduce the cutoff, consistent with \citet{Sironi25review}.

%%% ==================================================================================== %%%

\section{Conclusions and Future Prospects}
\label{sec:conclude}

We have investigated the role of guide fields on magnetic-energy dissipation, current-sheet dynamics, and nonthermal particle acceleration across a range of magnetisations in ion--electron semirelativistic reconnection. Our results demonstrate that the impact of the guide field is non-monotonic and it arises from a complex interplay between tearing, drift-kink, and MHD-like kink instabilities.

At low magnetisation, $\sigma_i= 0.1$, the evolution is largely tearing-dominated. Increasing the guide-field strength suppresses reconnection by suppressing the tearing modes and reducing the overall magnetic-energy dissipation. The drift-kink activity does not play a significant role; however, the sheet does kink at later times, leading to slower but additional magnetic-energy dissipation in comparison to equivalent 2D cases.

At high magnetisations ($\sigma_i= 1$ and $\sigma_i= 5$), the behaviour changes qualitatively. In the absence of a guide field, strong drift-kink activity corrugates and thickens the current sheet, inhibiting efficient tearing and reducing magnetic energy dissipation. Introducing a weak guide field suppresses this drift-kink-driven disruption, allowing the current sheet to remain thin and more coherent, thereby enhancing tearing-driven reconnection. The MHD-kink modes are not suppressed by a weak guide-field---they present another channel for energy loss. As the guide-field strength is further increased, reconnection is progressively suppressed: the onset is delayed, tearing and MHD-kink modes are suppressed, the active phase (plasma compression) weakens, and the system retains a larger fraction of its initial magnetic energy. The guide field plays a competing role: at low amplitudes it stabilises the current sheet against drift-kink-induced broadening, whilst at high amplitudes it suppresses reconnection by inhibiting compression and reducing the effective outflow speed. In essence, weak-to-moderate guide fields exacerbate reconnection (by suppressing the disruptive drift-kink modes), whilst strong guide fields suppress it.

Similar non-monotonic trends are seen for the particle energy distribution. At low magnetisation, reconnection primarily leads to plasma heating. At higher magnetisations, both electrons and ions develop suprathermal and nonthermal components. This nonthermal component is enhanced for weak-to-moderate guide fields compared to the zero-guide-field case, whereas strong guide fields steepen the spectra and reduce the high-energy cutoff by suppressing reconnection. A weak guide field may improve particle confinement and transport along the $z$-direction, allowing energetic particles to remain in the sheet for longer \citep{Dahlin2017}.

Our results indicate the presence of an optimal guide-field strength for which magnetic-energy dissipation and nonthermal particle acceleration are maximised. This threshold arises from the balance between suppressing drift-kink-induced current-sheet disruption and maintaining efficient tearing-driven reconnection. We infer that nonthermal acceleration in these simulations is controlled not only by the available magnetic energy but also by the morphology of the current sheet and the accessibility of coherent energising regions. In other words, the strongest accelerator is not necessarily the case with the weakest guide field, but rather the case in which the guide field is just strong enough to suppress drift-kink-driven broadening without yet quenching off reconnection.

The results presented here open several avenues for advancing our understanding of relativistic reconnection. A key priority is to establish a theory for the optimal guide-field strength as a function of magnetisation to maximise particle-acceleration efficiency. A useful conjecture is that this optimum occurs near the point where the guide-field-suppressed drift-kink growth rate becomes comparable or subdominant to the tearing growth rate,
\[ \Gamma_{\rm kink}(\sigma_i,B_z/B_0) \sim \Gamma_{\rm tear}(\sigma_i,B_z/B_0), \]
with the additional constraint that the guide-field-modified Alfv\'{e}n speed has not yet been substantially reduced. Establishing such a scaling, $B_{z,\rm opt}/B_0 = f(\sigma_i)$, will require a broader scan in $(\sigma_i,B_z/B_0)$ and measurements of the linear and nonlinear growth rates of the competing modes.

In the recent driven pair-plasma reconnection study of \citet{Camille26}, particle acceleration remained robust across changes in guide field and dimensionality, with asymptotic cutoffs converging to similar values and nonthermal slopes remaining in the range $s \simeq 1.6$--$2.0$. That result differs somewhat from the stronger guide-field dependence seen here, but the difference is physically understandable: their setup involves driven, strongly magnetised flux-tube merging in pair plasma, whereas ours probes semirelativistic ion--electron current-sheet reconnection in which drift-kink instability and ion--electron scale separation play a central role. In semirelativistic current sheets, as emphasised by \citet{Werner2024} and \citet{Bacchini2025}, the interplay between tearing, drift-kink, and flux-rope kinking can substantially alter both the dissipation history and the resulting particle acceleration.

In extreme compact-object environments, reconnection enters a high-energy-density regime in which radiative cooling, radiation pressure, Compton drag, and pair creation modify the thermodynamics and collisionality of the plasma \citep{Uzdensky2011}. \citet{Lyutikov2017a,Lyutikov2017b,Lyutikov2018} proposed that macroscopic magnetic stresses can instead drive explosive X-point collapse, producing charge-starved electric fields comparable to the reconnecting magnetic field and allowing particles to accelerate over macroscopic, rather than purely kinetic, length scales.

In subsequent work, we will move beyond idealised periodic setups toward more realistic configurations. In astrophysical systems, reconnection also occurs in dynamically evolving, turbulent environments rather than only in isolated current sheets. Embedding current sheets within self-consistent turbulent flows, or driving reconnection through large-scale forcing with nonperiodic boundaries, will allow us to determine whether the optimal-guide-field regime identified here persists under continuous energy injection and multiscale coupling. Such studies are expected to directly connect kinetic reconnection physics to global models of accretion flows, jets, and magnetically-dominated outflows.

Extending this framework to more realistic plasma conditions is essential for applicability in astrophysical context. This includes incorporating radiative cooling, anisotropic pressure effects, and pair versus ion--electron compositions. The goal is to connect the guide-field-regulated reconnection physics identified here to observable signatures, such as spectral slopes, variability timescales, and polarisation properties in high-energy astrophysical sources.

These directions point toward a comprehensive theory of reconnection in three dimensions, in which the guide field acts as a key regulator of the current-sheet structure, instability dynamics, and the efficiency of nonthermal particle acceleration.

%%% ==================================================================================== %%%

\section*{Acknowledgements}

PJD thanks Lorenzo Sironi, Luca Comisso, and Anatoly Spitkovsky for the stimulating discussions.
FB thanks Greg Werner and Dmitri Uzdensky for invaluable input and help throughout the development of this line of work.
PJD acknowledges support from the Research Foundation -- Flanders (FWO) via the long stay abroad grant V449025N. FB acknowledges support from the FED-tWIN programme (profile Prf-2020-004, project ``ENERGY''), issued by BELSPO, and from the FWO Junior Research Project G020224N granted by the Research Foundation -- Flanders (FWO). MZ acknowledges support from the NSF Grant PHY.~2512037.
The computational resources and services used in this work were provided, in part, by the VSC (Flemish Supercomputer Center), funded by the Research Foundation Flanders (FWO) and the Flemish Government – department WEWIS. We acknowledge LUMI-BE for awarding this project access to the LUMI supercomputer, owned by the EuroHPC Joint Undertaking, hosted by CSC (Finland) and the LUMI consortium through a LUMI-BE Regular Access call. We thank EuroHPC Joint Undertaking for awarding us access to the computational resources on the MeluXina (Luxembourg) and LUMI (Finland) supercomputers.
We used ChatGPT to develop Python codes for data analysis and visualisation of relevant data. 

%%%%%%%%%%%%%%%%%%%%%%%%%%%%%%%%%%%%%%%%%%%%%%%%%%
\section*{Data Availability}

The (relevant) data generated as part of this study is stored on LUMI, MeluXina, and VSC, and will be made available on reasonable request.



\bibliographystyle{mnras}
\bibliography{ref} % if your bibtex file is called example.bib

%\begin{thebibliography}{99}
%\end{thebibliography}

%%%%%%%%%%%%%%%%%%%%%%%%%%%%%%%%%%%%%%%%%%%%%%%%%%

%%%%%%%%%%%%%%%%% APPENDICES %%%%%%%%%%%%%%%%%%%%%

\appendix

\section{Size of the simulation domain}

\begin{figure*}
    \centering
    \includegraphics[width=\linewidth]{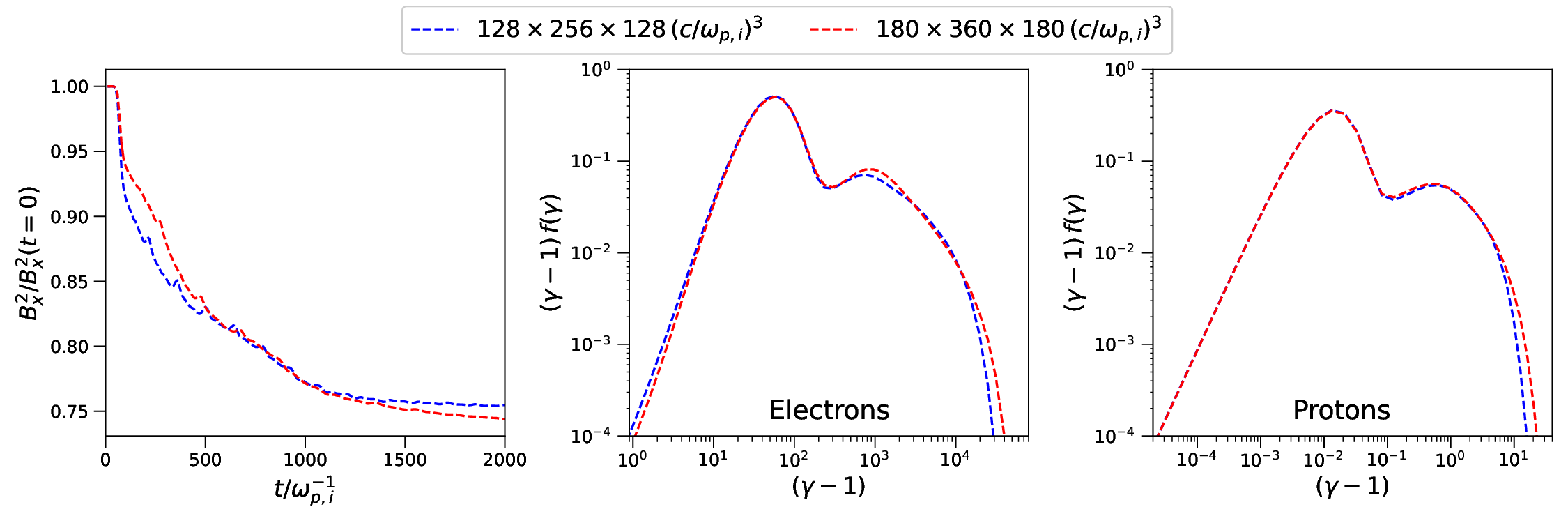}
    \caption{Comparison of two simulation domain sizes for the $\sigma_i=5$ and $B_z/B_0=0$ case. Left: temporal evolution of the reconnecting magnetic energy, $B_x^2/B_x^2(t=0)$. Middle: final electron energy spectrum. Right: final ion energy spectra. The blue dashed curves correspond to the domain $128\times256\times128\,(c/\omega_{p,i})^3$, and the red dashed curves to the larger domain $180\times360\times180\,(c/\omega_{p,i})^3$.}
    \label{fig:B_evolution_domain}
\end{figure*}

In order to investigate the effect of the simulation domain on the overall results, we present a comparison of a run with $\sigma_i=5$ with $B_z/B_0 = 0$ with two different simulation domains (Fig.~\ref{fig:B_evolution_domain}). The larger system size of $180 \times 360 \times 180 \, (c/\omega_{p, i})^3$ was the size chosen by \citet{Bacchini2025} for their simulation run with $\sigma_i= 10$. Any quantitative differences in magnetic-energy dissipation and particle energy spectra are minor; the qualitative behaviour is consistent. This validates our justification for choosing system sizes based on the upstream Larmor radius for varying magnetisations.

% \section{FFT of tearing and drift-kink modes}

% Fig. \ref{fig:kink_modes} shows ...

% $k_{min}$ is set by the domain size; $k_{max}$ is set by the grid resolution; number of k modes is limited by the resolution and system size - too coarse grids cannot resolve small scale structures, i.e., energy in the small wavenumbers

%%%%%%%%%%%%%%%%%%%%%%%%%%%%%%%%%%%%%%%%%%%%%%%%%%

% Don't change these lines
\bsp	% typesetting comment
\label{lastpage}
\end{document}